\documentclass[aps,prd,reprint,groupedaddress]{revtex4-2}
\usepackage{amsmath}
\usepackage{amsfonts}
\usepackage{mathrsfs}
\usepackage{amssymb}
\usepackage{bm}

\newcommand{\normord}[1]{\;\bm{\vcentcolon}\,\mathrel{#1}\,\bm{\vcentcolon}\;}
\providecommand{\vcentcolon}{\mathrel{\mathop{:}}} 

\usepackage{slashed}

\begin{document}

%\preprint{}

\title{Radiative corrections in the Yukawa model\\ within the null-plane causal perturbation theory framework}


\author{O.A. Acevedo}
\email[]{oscar.acevedo@unesp.br}
%\thanks{}
%\affiliation{}

\author{B.M. Pimentel}
\email[]{bruto.max@unesp.br}
%\thanks{}
\affiliation{Institute for Theoretical Physics, São Paulo State University, Brazil}


\date{\today}

\begin{abstract}
Practical calculations in light-front dynamics are, as a general rule, complicated, since there is no consensus about how to treat the poles which come from the instantaneous parts of Feynman's propagators of the fields. An alternative to solve this difficulty is null-plane causal perturbation theory, a recent developed framework which prevents the appearance of the mentioned poles by avoiding the usage of Feynman's propagators in ``loop distributions'', requiring no regularization of the amplitudes. In this study, we treat the radiative corrections in the neutral Yukawa's model in that framework. Particularly, we explicitly calculate the boson and fermion self-energies and show that the results obtained with this approach are equivalent to that of the instant dynamics.
\end{abstract}

%\keywords{}

\maketitle


\section{Introduction}

As is well known, Yukawa's model \cite{Yukawa1935, Kemmer1938, Wick1938, Kemmer1938b} is a phenomenological model for the interaction between nucleons and pions. It was studied in light-front dynamics by Chang and Yan \cite{ChangYan} in 1973 by using Schwinger's functional derivative method. The problem of finding the fermion self-energy for an analogous model, with scalar instead of pseudo-scalar mesons\footnote{Such a model, a little farther from the real situation --the pions are pseudo-scalar particles--, is commonly also called Yukawa's model. It differs from the one considered here in very little aspects: The integrals needed for the calculation of radiative corrections are the same in both cases, but enter into the $S$-matrix with different numerical coefficients.}, in light-front dynamics was also recently considered by Karmanov, Mathiot and Smirnov \cite{Karmanov}; these investigators pointed out some difficulties: (i) the cut-offs of the null plane variables lead to a dependence of the amplitudes on the null plane orientation in some regularization techniques; (ii) such a dependence is very sensitive because each partial amplitude in light-front dynamics usually diverges more strongly than the corresponding Feynman's amplitude; (iii) that dependence disappears in the re-normalized amplitudes, but the re-normalization procedures are drastically different for different regularization schemes. Also, Bakker and Ji \cite{Bakker} have considered box diagrams for light-front (scalar) Yukawa's model; they conclude that the question of what regularization scheme is the one which would lead to an invariant $S$-matrix remains, to the date, without an answer. Another study in this problem was done by Grangé, Mathiot, Mutet and Werner \citep{Grange1,Grange2,Grange3}, which considered the distributional character of the quantized fields and used Taylor-Lagrange's regularization in order to make finite the Feynman's amplitudes.

Recently, the authors have developed causal perturbation theory (CPT) on light-front dynamics \cite{APS, APS2}, a perturbation solution to Bogoliubov-Medvedev-Polivanov's axioms \cite{BogoMedPoli, Bogo, BogoLogunov} for Heisenberg's $S$-matrix program \cite{HeisenbergS} which is ultra-violet finite and aims to clarify, by introducing from the very beginning the causality axiom, referred to the $x^+$ time coordinate, and performing --following Epstein and Glaser \cite{EpsteinGlaser}-- well-defined distributional operations only, the true meaning of the instantaneous terms which appear both in the propagators and Hamiltonian in the usual formulation of light-front QFT --see, for example, Ref. \cite{Brodsky1}--. This approach is inspired in the analogous instant dynamics formulation of CPT \cite{EpsteinGlaser,ScharfFQED,ScharfGFT,Prange,Grigore,Duetsch}. In this paper we re-examine the radiative corrections for the neutral Yukawa's model \cite{Bogo}, which in the usual approach lead to the already commented ambiguities regarding the regularization processes. In the causal approach, being finite by construction, those operations are not needed; therefore, since they are identified as the origin of some problems, we hope that null plane CPT could make simpler the comparison of light-front QFT with the usual instant dynamics formulation. Additionally, in Ref. \cite{APS2} it was shown that the singular order of every causal distribution in null plane Yukawa's model is the same as in instant dynamics, and, accordingly, the normalization procedure in this theoretical framework is not more difficult, but, in fact, it is the same. Also regarding the invariance of the $S$-matrix to all orders null plane CPT can shed some light: As an inductive theory, each causal distribution at each order is constructed with the previous transition distributions; since these must be already normalized, the instantaneous terms do not propagate to the following orders. In order to familiarize the reader with null plane CPT we present a very brief résumé at next --the details of the construction of the theory can be found in Ref. \cite{APS2}--.

\subsection{Null-plane causal perturbation theory}\label{subsec:I.A}

CPT uses the operation of ``adiabatic switching'' \cite{Bogo}, which consists in multiply the coupling constant of the interaction theory by a ``switching function'' $g\in \mathscr S(\mathbb{R}^4):\mathbb{R}^4\to\mathbb{R}$, in order to isolate the problem of infra-red divergences; for Yukawa's model, in which the meson is massive, there are no infra-red divergences and the adiabatic limit $g\to 1$, by means of which the real interaction is recovered, is trivial.

The general properties that the scattering operator $S(g)$ must respect constitute the Bogoliubov-Medvedev-Polivanov axioms \cite{BogoMedPoli, Bogo, BogoLogunov}; however, following Epstein and Glaser \cite{EpsteinGlaser}, we only consider the axioms which will be used for the construction of CPT: (i) translation invariance and (ii) causality --now referred to the $x^+$ null plane time--. The remaining axioms are not needed for the theoretical construction, but will be considered as physical conditions for the normalization of the solutions; they are (iii) unitarity, (iv) Lorentz's invariance and (v) when possible, vacuum stability, which in CPT is related to the problem of the adiabatic limit.

CPT is a perturbation theory; therefore the $S(g)$ operator is written as the following series:
\begin{equation}\label{eq:3.2.3}
S(g)=1+\sum\limits_{n=1}^{+\infty}\frac{1}{n!}\int dXT_n(X)g(X)\ ,
\end{equation}
with $X:=\left\lbrace x_j\in\mathbb{M}\ |\ j=1,\cdots,n\right\rbrace$ and the notations:
\begin{equation*}
T_n(X)\equiv T_n(x_1;\cdots;x_n)\ ,\  g(X)\equiv g(x_1)\cdots g(x_n)\ ,
\end{equation*}
\begin{equation}
dX\equiv d^4x_1\cdots d^4x_n\ .\label{eq:3.2.2}
\end{equation}
Eq. \eqref{eq:3.2.3} serves as the definition of the operator-valued distributions $T_n(x_1;\cdots;x_n)\in\mathscr S'(\mathbb{R}^{4n})$, called ``transition distributions of order $n$'' or ``$n$-point distributions''. 

The inverse operator $S(g)^{-1}$ is given as a perturbation series as well:
\begin{equation*}
S(g)^{-1}=1+\sum\limits_{n=1}^{+\infty}\frac{1}{n!}\int dX\widetilde T_n(X)g(X)\ ; 
\end{equation*}
\begin{equation}\label{eq:3.2.4}
\widetilde T_n(X)=\sum\limits_{r=1}^n(-1)^r\sum\limits_{\begin{subarray}{l} X_1,\cdots,X_r\neq\emptyset\\ X_1\cup\cdots\cup X_r=X\\ X_j\cap X_k=\emptyset, \forall j\neq k \end{subarray}}T_{n_1}(X_1)\cdots T_{n_r}(X_r)\ .
\end{equation}

The causality axiom implies that the transition distributions are ``chronologically ordered'':
\begin{equation}\label{eq:3.2.17}
T_n(X)=T_m(X_2)T_{n-m}(X_1)\ \text{for}\  X_1<X_2\ ; 
\end{equation}
\begin{equation}\label{eq:3.2.17-1}
\left[T_n(X);T_m(Y)\right]=0\ \text{for}\  X\sim Y\ .
\end{equation}
This result as a consequence of the causality axiom was first established by Bogoliubov and Shirkov \cite{Bogo}. However, they failed in the way of chronologically order these products: They took the product by Heaviside's functions in order to make contact with the more usual formulation of QFT by Feynman and others, and then arrived to the same ultra-violet divergences. To see how extremely important is to take care in performing such multiplications, let us consider as an example the distribution $\delta(x)$. Its product by a Heaviside's function $\Theta(x)$ has Fourier's transform:
\begin{align*}
\widehat{\Theta\delta}(p)&=(2\pi)^{-1/2}\int dq\widehat\Theta(q)\hat\delta(p-q)\\
&=i(2\pi)^{-3/2}\int\frac{dq}{q+i0^+}\ ,
\end{align*}
which is ultra-violet divergent. It was Bogoliubov and Parasiuk \cite{BogoParasiuk} who discover that the ultra-violet divergences in QFT are due to the presence of products of distributions with discontinuous functions as Heaviside's. Epstein and Glaser, then, modified the prescription of chronological order: To obtain the transition distribution they did not use that ill-defined recipe, but the causality axiom to perform the following (distributional) well-defined construction: Define the ``advanced distribution of order $n$'' as the following distribution:
\begin{equation}\label{eq:3.3.2}
A_n(Y;x_n)=\sum\limits_{\begin{subarray}{l} X\cup X'=Y\\ X\cap X'=\emptyset\end{subarray}}\widetilde T_m(X)T_{n-m}(X'\cup\left\{x_n\right\})\ ,
\end{equation}
and the ``retarded distribution of order $n$'' as:
\begin{equation}\label{eq:3.3.4}
R_n(Y;x_n)=\sum\limits_{\begin{subarray}{l} X\cup X'=Y\\ X\cap X'=\emptyset\end{subarray}}T_{n-m}(X'\cup\left\{x_n\right\})\widetilde T_m(X)\ .
\end{equation}

In the sums in Eqs. \eqref{eq:3.3.2} and \eqref{eq:3.3.4} the $n$-point distribution appears once. Separating it from the other terms:
\begin{equation}\label{eq:3.3.5}
A_n(Y;x_n)=T_n(Y\cup\left\{x_n\right\})+A'_n(Y;x_n)\ ,
\end{equation}
\begin{equation}\label{eq:3.3.5.1}
R_n(Y;x_n)=T_n(Y\cup\left\{x_n\right\})+R'_n(Y;x_n)\ ,
\end{equation}
with the following definitions of the ``advanced subsidiary distribution'' and of the ``retarded subsidiary distribution'', respectively:
\begin{equation}\label{eq:3.3.6}
A'_n(Y;x_n):=\sum\limits_{\begin{subarray}{l} X\cup X'=Y\\ X\cap X'=\emptyset\\ X\neq\emptyset\end{subarray}}\widetilde T_m(X)T_{n-m}(X'\cup\left\{x_n\right\})\ ,
\end{equation}
\begin{equation}\label{eq:3.3.7}
R'_n(Y;x_n):=\sum\limits_{\begin{subarray}{l} X\cup X'=Y\\ X\cap X'=\emptyset\\ X\neq\emptyset\end{subarray}}T_{n-m}(X'\cup\left\{x_n\right\})\widetilde T_m(X)\ .
\end{equation}
These subsidiary distributions do not contain $T_n$, but only the transition distributions $T_m$ with $m\leq n-1$. The transition distribution of order $n$ is then equal to:
\begin{align}
T_n(Y\cup\left\{x_n\right\})&=A_n(Y;x_n)-A'_n(Y;x_n)\nonumber\\
&=R_n(Y;x_n)-R'_n(Y;x_n)\ .\label{eq:3.3.8}
\end{align}
In this way, the $n$-point distribution can be found if we know the distributions $T_m$ with $m\leq n-1$ --this is the inductive hypothesis-- and the advanced or retarded distribution of order $n$, which can be found by splitting \cite{Division1, Division2, Division3} the ``causal distribution of order $n$'':
\begin{align}
D_n(Y;x_n):&=R_n(Y;x_n)-A_n(Y;x_n)\nonumber\\
&=R'_n(Y;x_n)-A'_n(Y;x_n)\ ,\label{eq:3.3.9}
\end{align}
which is known once the subsidiary distributions are. The splitting procedure is based on the results: (i) the support of the retarded (resp. advanced) distribution is contained in\footnote{By definition, $\Gamma_n^{\pm}(x)$ is the set consisting of all $n$-tuples of points in Minkowski's space-time which are in $V^\pm\cup x^-$-axis, with respect to the point $x$.} $\Gamma_n^\pm(x_n)$, and (ii) the support of $D_n$ is causal for $n\geq 3$, while for $n=2$ it must be proven in each particular model. It must be done as follows: Suppose that the causal distribution of order $n$ was already constructed by means of the inductive procedure, and that it has causal support. In general, it will have the following form:
\begin{equation}\label{eq:3.5.1}
D_n(x_1;\cdots;x_n)=\sum_kd^k_n(x_1;\cdots;x_n)\normord{C_k(u^A)}\ ,
\end{equation}
with $d^k_n(x_1;\cdots;x_n)$ a numerical distribution and $\normord{C_k(u^A)}$ a Wick's monomial of the different quantized free operator fields $u^A$ of the theory. Since this operator fields do not restrict the support of the complete distribution --they are defined in all Minkowski's space-time--, it is sufficient to consider the split of the numerical distribution $d^k_n$, whose support, then, is causal by hypothesis. Also the advanced and retarded distributions will maintain the operator fields structure of the causal distribution:
\begin{equation}\label{eq:3.5.2}
A_n(x_1;\cdots;x_n)=\sum_ka^k_n(x_1;\cdots;x_n)\normord{C_k(u^A)}\ ,
\end{equation}
\begin{equation}\label{eq:3.5.3}
R_n(x_1;\cdots;x_n)=\sum_kr^k_n(x_1;\cdots;x_n)\normord{C_k(u^A)}\ ,
\end{equation}
with $a^k_n$ and $r^k_n$ the advanced and retarded parts, respectively, of the numerical distribution $d^k_n$. Using the translational invariance, define the numerical distribution $d\in\mathscr{S}(\mathbb{R}^{4n-4})$ as:
\begin{equation}\label{eq:3.5.6}
d(x):=d^k_n(x_1-x_n;\cdots;x_{n-1}-x_n;0)\ ; 
\end{equation}
\begin{equation}\label{eq:3.5.6-1}
\text{supp}(d)\subseteq\Gamma^+_{n-1}(0)\cup\Gamma^-_{n-1}(0)\ ,
\end{equation}
which will be split as:
\begin{equation}\label{eq:3.5.7}
d=r-a\ ;\ \text{supp}(r)\subseteq\Gamma^+_{n-1}(0)\ ,\ \text{supp}(a)\subseteq\Gamma^-_{n-1}(0)\ .
\end{equation}
In Eq. \eqref{eq:3.5.6} we have written $d(x)$; $x$ means: $(x_1-x_n;\cdots;x_{n-1}-x_n)$. In the following we will use Schwartz's multi-index notation\footnote{It is defined, for example, in Ref. \cite{EpsteinGlaser}: A multi-index $k\in\mathbb{R}^N$ is a sequence of non-negative numbers, $k=(k_1;\cdots;k_N)$, $k_j\geq0$, for which the following notations are established:
\begin{equation*}
|k|\equiv\sum\limits_{j=1}^N k_j\ ,\  x^k\equiv\prod\limits_{j=1}^Nx_j^{k_j}\ ,\ k!\equiv\prod\limits_{j=1}^Nk_j!\ ,
\end{equation*}
\begin{equation*}
 D^kf(x)\equiv\prod\limits_{j=1}^N\partial_{x_j}^{k_j}f(x)\ .
\end{equation*}} We will also use the notation: $x^a\equiv(x_1^a-x_n^a;\cdots;x_{n-1}^a-x_n^a)$.

In order to perform the splitting, the first natural attempt is to multiply by Heaviside's functions. But, as we already know, such a multiplication is ill-defined if the singularities of the distribution lay on the discontinuity surface of the function --as it was exemplified by the product $\Theta(x)\delta(x)$--; this is the second important contribution of Epstein and Glaser \cite{EpsteinGlaser}: they use the distribution splitting theory \cite{Division1, Division2, Division3} to perform those operations. As it was said, to do that it is very important to know how the distribution behaves near the splitting point. Fassari and Scharf \cite{Fassari} simplified the analysis of the singularity of distributions made by Epstein and Glaser by using the concept of quasi-asymptotics developed by Vladimirov, Drozzinov and Zavialov \cite{Vladimirov}. In null-plane dynamics, since the planes of constant $x^+$ time intersects the light-cone all long way the $x^-$-axis, it is the behaviour of the distribution $d(x)$ near the $x^-$-axis that is essential for the splitting procedure. Such behaviour will be examined at the light of the following definition of quasi-asymptotics by selected variable \cite{Vladimirov}:

\textbf{Definition:} \textit{Let $d\in\mathscr{S}'(\mathbb{R}^m)$ be a distribution, and let $\rho$ be a continuous positive function. If the (distributional) limit}
\begin{equation}\label{eq:3.5.9}
\lim_{s\to 0^+}\rho(s)s^{3m/4}d\left(sx^+;sx^\perp;x^-\right)=d_-(x)
\end{equation}
\textit{exists in $\mathscr{S}'(\mathbb{R}^m)$ and is non-null, then the distribution $d_-$ is called the \underline{quasi-asymptotics} of $d$ at the $x^-$-axis, with regard to the function $\rho$.} $\square$

With this definition, the function $\rho(s)$ can be shown to be a ``regularly varying at zero function'', also called an ``auto-model function'' \cite{Vladimirov, ScharfFQED}, which means that for every $a>0$:
\begin{equation}\label{eq:3.5.9.1}
\lim_{s\to0^+}\frac{\rho(as)}{\rho(s)}=a^\alpha\ ,
\end{equation}
for some $\alpha\in\mathbb{R}$, called the ``order of auto-modelity'' of the function $\rho$. This number serves as a characterizing parameter of the distribution, which is called its ``singular order at the $x^-$-axis'' and is denoted by $\omega_-$. The singular order of the causal distribution at the $x^-$-axis is extremely important, as it determines the space of test functions on which the retarded distribution can be defined in principle: (1) For negative singular order, $\omega_-<0$, it is the entire Schwartz's space $\mathscr S$, and the retarded distribution can be obtained by simple multiplication of the causal distribution by Heaviside's function. In momentum space the following splitting formula holds for this case:
\begin{align}
\hat r(p)&=\frac{i}{2\pi}\int\limits_{-\infty}^{+\infty}\frac{\hat d\left((p_{1+}-k;\bm p_1);\cdots;(p_{n-1+}-k;\bm p_{n-1})\right)}{k+i0^+}dk\nonumber\\
&\equiv\frac{i}{2\pi}\int\limits_{-\infty}^{+\infty} \frac{\hat d(p_+-k;\bm p)}{k+i0^+}dk\ .\label{eq:3.6.9}
\end{align}
(2) For non-negative singular order, $\omega_-\geq0$, the retarded distribution can only be defined (in principle) on the space of test functions for which the first $\omega_-$ derivatives at the $x^-$-axis vanish. Such a retarded distribution can be extended to the whole $\mathscr S$ by means of the so-called $W$-operation, which projects the general test function onto the restricted space in which the retarded distribution is defined. By transposing the $W$-operation to the distribution and going to momentum space, we obtain the ``retarded distribution with normalization line $\left(q_+;q_\perp;p_-\right)$'':
\begin{align}
&\hat r_q(p)=\frac{i}{2\pi}\int\limits_{-\infty}^{+\infty}\frac{dk}{k+i0^+}\Bigg\{\hat d(p_+-k;\bm p)\nonumber\\
&-\sum\limits_{|c|=0}^{\lfloor\omega_-\rfloor}\frac{1}{c!}(p_{+,\alpha}-q_{+,\alpha})^cD^c_{+,\alpha}\hat d(q_+-k;q_\perp;p_-)\Bigg\}\ ,\label{eq:3.6.47}
\end{align}
which is a well-defined distribution on the entire $\mathscr S$. A particular case of normalization line is $(0;0_\perp;p_-)$; the solution normalized at it is called the ``central solution'', and is the one we will use in this paper.

Finally, if $(r_1,a_1)$ and $(r_2,a_2)$ are two solutions of the splitting problem, then by Eq. \eqref{eq:3.3.9} we have that $r_1-a_1=r_2-a_2$, so that $r_1-r_2=a_1-a_2$. Since the left-hand-side of this equation has support on $\Gamma^+$, while its right-hand-side has support in $\Gamma^-$, the difference $(r_1-r_2)$ can only have support on $\Gamma^+\cap\Gamma^-=x^-$-axis. Therefore, $r_1$ and $r_2$ could be different only by ``normalization terms'' which are distributions with support on the $x^-$-axis. In momentum space:
\begin{equation}\label{eq:3.7.5}
\hat r_1(p)-\hat r_2(p)=\sum\limits_{|b|=0}^{M}\widehat C_b\left(p_-\right)p_{+,\perp}^b\ ,
\end{equation}
with $\widehat C_b\left(p_-\right)$ some distributions of the variable $p_-$. The singular order of each one of these terms is $|b|$, independently of which the distribution $\widehat C_b(p_-)$ is, because the variable $p_-$ is not scaled in the singular order calculus. The procedure of determining these unknown distributions by the imposition of physical requirements is called the ``normalization process''.

\section{Meson's self-energy}

Yukawa's model is defined by the one-point distribution
\begin{equation}\label{eq:4.2.0}
T_1(x)=-i\mathfrak{g}\normord{\overline\psi(x)\gamma^5\psi(x)}\varphi(x)\ .
\end{equation}
According to the inductive procedure [Eqs. \eqref{eq:3.3.6}, \eqref{eq:3.3.7} and \eqref{eq:3.3.9}], the causal distribution of the second order is obtained from $T_1$ as:
\begin{equation}\label{eq:4.2.0-1}
D_2(x_1;x_2)=\left[T_1(x_1);T_1(x_2)\right]\ .
\end{equation}
Replacing here the distribution from Eq. \eqref{eq:4.2.0} and using Wick's theorem, it is shown that the second order causal distribution, describing meson's self-energy, is given by \cite{APS2} --we use in the following the relative coordinate $y=x_1-x_2$--:
\begin{equation}\label{eq:4.2.1}
D_2^{(BSE)}(x_1;x_2)=\normord{\varphi(x_1)d(y)\varphi(x_2)}\ ,
\end{equation}
with the following definitions:
\begin{equation}\label{eq:4.2.2}
d(y):=P(y)-P(-y)\ ; 
\end{equation}
\begin{equation*}
P(y):=\mathfrak g^2\text{Tr}\left[S_+(y)\gamma^5 S_-(-y)\gamma^5\right]\ .
\end{equation*}
Already at this point we can see one very important advantage of null-plane CPT: Different from other approaches which use Feynman's rules, Feynman's propagators do not appear in our ``loop distributions'', so the instantaneous term in the fermion propagator, which contains a spurious pole whose removal has been matter of study for many years \cite{Karmanov, Bakker,Grange3}, simply does not appear.

In order to go to momentum space we apply Fourier's transformation to $P(y)$, obtaining:
\begin{align}
\widehat P(q)&=(2\pi)^{-2}\mathfrak g^2\int d^4p\text{Tr}\left[\widehat S_+(p)\gamma^5\widehat S_-(p-q)\gamma^5\right]\nonumber\\
&=(2\pi)^{-2}\mathfrak g^2\int d^4p\text{Tr}\left[(\slashed p+m_1)\gamma^5(\slashed p-\slashed q+m_1)\gamma^5\right]\nonumber\\
&\quad\times\widehat D_+(p)\widehat D_-(p-q)\ .\label{eq:4.2.3}
\end{align}
Using that $\gamma^\mu\gamma^5=-\gamma^5\gamma^\mu$, the trace in Eq. \eqref{eq:4.2.3} is:
\begin{equation}\label{eq:4.2.4}
\text{Tr}\left[(\slashed p+m_1)\gamma^5(\slashed p-\slashed q+m_1)\gamma^5\right]=4\left(m_1^2-p^2+pq\right)\ ,
\end{equation}
so that, remembering that 
\begin{equation*}
\widehat D_\pm(p)=\pm i(2\pi)^{-1}\Theta\left(\pm p_-\right)\delta(p^2-m_1^2)\ ,
\end{equation*}
we obtain:
\begin{align}\label{eq:4.2.5}
\widehat P(q)=&4(2\pi)^{-4}\mathfrak g^2\int d^4p\left(m_1^2-p^2+pq\right)\Theta\left(p_-\right)\nonumber\\
&\times\Theta\left(q_--p_-\right)\delta(p^2-m_1^2)\delta\left((p-q)^2-m_1^2\right)\ .
\end{align}
The supports of Dirac's delta distributions appearing here are:
\begin{equation}\label{eq:4.2.6}
p^2=m_1^2\quad\text{and}\quad q^2=2pq\ .
\end{equation}
Therefore,  Eq. \eqref{eq:4.2.5} is equal to:
\begin{equation}\label{eq:4.2.7}
\widehat P(q)=2(2\pi)^{-4}\mathfrak g^2q^2I(q)\ ,
\end{equation}
with:
\begin{equation}\label{eq:4.2.8}
I(q)=\int d^4p\Theta\left(p_-\right)\Theta\left(q_--p_-\right)\delta(p^2-m_1^2)\delta(q^2-2pq)\ .
\end{equation}
In order to calculate this integral we move to an appropriate reference frame. Since $p, q-p\in V^+$, then also $q=p+(q-p)\in V^+$, and there is a reference frame in which $q=(q_+;0_\perp;q_-)$. In that reference frame:
\begin{align}
I(q)&=\int d^4p\Theta\left(p_-\right)\Theta\left(q_--p_-\right)\frac{1}{\left|2p_-\right|}\delta\left(p_+-\frac{\omega_p^2}{2p_-}\right)\nonumber\\
&\quad\times\delta\left(2q_+q_--2q_-p_+-2q_+p_-\right)\nonumber\\
&=\int dp_-d^2p_\perp\Theta\left(p_-\right)\Theta\left(q_--p_-\right)\frac{1}{|4q_+|}\nonumber\\
&\quad\times\delta\left(\left(p_-\frac{q_-}{2}\right)^2-A^2\right)\ ,\label{eq:4.2.9}
\end{align}
with:
\begin{equation}\label{eq:4.2.10}
A:=\sqrt{\frac{q_-^2}{4}-\frac{q_-\omega_p^2}{2q_+}}\ .
\end{equation}
Performing the integration:
\begin{align}
I(q)&=\int dp_-d^2p_\perp\frac{\Theta\left(p_-\right)\Theta\left(q_--p_-\right)}{|4q_+||2A|}\nonumber\\
&\quad\times\left\lbrace\delta\left(p_--\left[\frac{q_-}{2}-A\right]\right)+\delta\left(p_--\left[\frac{q_-}{2}+A\right]\right)\right\rbrace\nonumber\\
&=\frac{\pi}{|4q_+|}\Theta(q_-)\Theta(2q_+q_--4m_1^2)\int\limits_{m_1^2}^{q_+q_-/2}\frac{d(\omega_p^2)}{\sqrt{\frac{q_-^2}{4}-\frac{q_-\omega_p^2}{2q_+}}}\nonumber\\
&=\frac{\pi}{2}\Theta(q_-)\Theta(2q_+q_--4m_1^2)\sqrt{1-\frac{4m_1^2}{2q_+q_-}}\ .\label{eq:4.2.11}
\end{align}
In Lorentz's covariant form this integral is:
\begin{equation}\label{eq:4.2.12}
I(q)=\frac{\pi}{2}\Theta\left(q_-\right)\Theta(q^2-4m_1^2)\sqrt{1-\frac{4m_1^2}{q^2}}\ ,
\end{equation}
and, substituting into Eq. \eqref{eq:4.2.7}:
\begin{equation}\label{eq:4.2.13}
\widehat P(q)=\frac{1}{2}(2\pi)^{-3}\mathfrak g^2q^2\Theta\left(q_-\right)\Theta(q^2-4m_1^2)\sqrt{1-\frac{4m_1^2}{q^2}}\ .
\end{equation}
Accordingly, the numerical part of the causal distribution for boson's self-energy has Fourier's transform [see Eq. \eqref{eq:4.2.2}]:
\begin{equation}\label{eq:4.2.14}
\hat d(q)=\frac{1}{2}(2\pi)^{-3}\mathfrak g^2q^2\text{sgn}\left(q_-\right)\Theta(q^2-4m_1^2)\sqrt{1-\frac{4m_1^2}{q^2}}\ .
\end{equation}

By the supports of Heaviside's function and Dirac's delta distribution we have that $\hat d(q)$ is non-null only when $q\in V^+$; therefore there exists a reference frame in which $(q_a)=(q_+;0_\perp;q_-)$; in that reference frame:
\begin{align}
\hat d(q)=&(2\pi)^{-3}\mathfrak g^2q_+q_-\text{sgn}\left(q_-\right)\Theta(2q_+q_--4m_1^2)\nonumber\\
&\times\sqrt{1-\frac{4m_1^2}{2q_+q_-}}\ .\label{eq:4.2.15}
\end{align}
According to the factorization of polynomials theorem --see Ref. \cite{Duetsch}--, it will be sufficient to split the distribution
\begin{equation}\label{eq:4.2.16}
\hat d_1(q)=\text{sgn}\left(q_-\right)\Theta(2q_+q_--4m_1^2)\sqrt{1-\frac{4m_1^2}{2q_+q_-}}\ ,
\end{equation}
whose singular order at the $x^-$-axis is:
\begin{equation}\label{eq:4.2.17}
\omega_-^1=0\ ,
\end{equation}
so that its retarded part, according to Eq. \eqref{eq:3.6.47} and choosing the central solution, is obtained as:
\begin{align}
\hat r_1(q)=&\frac{i}{2\pi}\int\frac{dk}{k+i0^+}\big\{\hat d_1\left(q_+-k;0_\perp;q_-\right)\nonumber\\
&-\hat d_1\left(-k;0_\perp;q_-\right)\big\}\ .\label{eq:4.2.18}
\end{align}
Substituting Eq. \eqref{eq:4.2.16} into Eq. \eqref{eq:4.2.18} we will have, using the variable $s=-2q_-k$:
\begin{align}
\hat r_1(q)&=-\frac{i}{2\pi}\int \frac{ds}{s-iq_-0^+}\Bigg\{\Theta(s+2q_+q_--4m_1^2)\nonumber\\
&\times\sqrt{1-\frac{4m_1^2}{s+2q_+q_-}}-\Theta(s-4m_1^2)\sqrt{1-\frac{4m_1^2}{s}}\Bigg\}\ .\label{eq:4.2.19}
\end{align}
Applying Sokhotskiy's formula into the first integral, then making $s+2q_+q_-\to s$ in it and joining with the second term in Eq. \eqref{eq:4.2.19}, we obtain:
\begin{align}
\hat r_1(q)=&-\frac{i}{2\pi}2q_+q_-\int\limits_{4m_1^2}^{+\infty}\frac{ds}{s(s-2q_+q_-)}\sqrt{1-\frac{4m_1^2}{s}}\nonumber\\
&+\frac{1}{2}\text{sgn}\left(q_-\right)\Theta(2q_+q_--4m_1^2)\sqrt{1-\frac{4m_1^2}{2q_+q_-}}\ .\label{eq:4.2.20}
\end{align}
The remaining integral can be found by using Euler's substitution:
\begin{equation}\label{eq:4.2.21}
 \frac{s}{m_1^2}=\frac{(1+x)^2}{x}\quad(0<x<1)\ ; \frac{ds}{m_1^2}=-\frac{1-x^2}{x^2}dx\ ;
\end{equation}
\begin{equation*}
\sqrt{1-\frac{4m_1^2}{s}}=\frac{1-x}{1+x}\ .
\end{equation*}
We also define the parameter $\xi$ as:
\begin{equation}\label{eq:4.2.22}
\frac{2q_+q_-}{m_1^2}=-\frac{(1-\xi)^2}{\xi}\ ;\  \xi=\frac{\sqrt{1-4m_1^2/2q_+q_-}-1}{\sqrt{1-4m_1^2/2q_+q_-}+1}\ .
\end{equation}
Therefore:
\begin{align}\label{eq:4.2.23}
&\int\limits_{4m_1^2}^{+\infty}\frac{ds}{s(s-2q_+q_-)}\sqrt{1-\frac{4m_1^2}{s}}\nonumber\\
&\quad=\frac{1}{m_1^2}\int\limits_0^1\frac{dx(1-x)^2}{(1+x)^2(x+\xi)(x+1/\xi)}\ ,
\end{align}
which, having rational integrand, can already be calculated by the technique of partial fractions decomposition. It is found that Eq. \eqref{eq:4.2.23} is equal to:
\begin{equation}\label{eq:4.2.24}
-\frac{1}{m_1^2}\frac{1}{(1-\xi)^3}\left[\xi(1+\xi)\log(|\xi|)+2\xi(1-\xi)\right]\ ,
\end{equation}
whose substitution into Eq. \eqref{eq:4.2.20} leads us to:
\begin{align}
\hat r_1(q)&=-\frac{i}{2\pi}\left[\frac{1+\xi}{1-\xi}\log(|\xi|)+2\right]\nonumber\\
&+\frac{1}{2}\text{sgn}\left(q_-\right)\Theta(2q_+q_--4m_1^2)\sqrt{1-\frac{4m_1^2}{2q_+q_-}}\ .\label{eq:4.2.25}
\end{align}

The retarded distribution corresponding to the distribution $\hat d(q)$ in Eq. \eqref{eq:4.2.15} is therefore, already in Lorentz's covariant form --which is obtained by the replacement of $2q_+q_-$ by $q^2$--: 
\begin{align}
\hat r(q)=&-\frac{i\mathfrak g^2}{2(2\pi)^4}q^2\Bigg\{\frac{1+\xi}{1-\xi}\log(|\xi|)+2\nonumber\\
&-i\pi\Theta(q^2-4m_1^2)\sqrt{1-\frac{4m_1^2}{q^2}}\Bigg\}\ .\label{eq:4.2.27}
\end{align}
Subtracting the subsidiary distribution $\hat r'(q)$ --whose only effect is to change $\text{sgn}\left(q_-\right)$ by $1$ in the imaginary term coming from Sokhotskiy's formula-- to obtain the distribution $\hat t(q)=\hat r(q)-\hat r'(q)$, and defining:
\begin{equation*}
\hat t(q):=-i\widehat\Pi(q)\ ,  
\end{equation*}
\begin{equation}\label{eq:4.2.28}
T_2^{(BSE)}(x_1;x_2)=-i\normord{\varphi(x_1)\Pi(x_1-x_2)\varphi(x_2)}\ ,
\end{equation}
with $\Pi$ being the so-called ``boson's self-energy'', we arrive at the result:
\begin{align}
\widehat \Pi(q)=&\frac{\mathfrak g^2}{2(2\pi)^4}q^2\Bigg\{\frac{1+\xi}{1-\xi}\log(|\xi|)+2\nonumber\\
&-i\pi\Theta(q^2-4m_1^2)\sqrt{1-\frac{4m_1^2}{q^2}}\Bigg\}\ .\label{eq:4.2.29}
\end{align}

Now, the whole $\widehat\Pi(q)$ distribution has singular order $\omega_-=+2$, so its most general form is:
\begin{equation}\label{eq:4.2.30}
\widetilde\Pi(q)=\widehat\Pi(q)+C_0+c_aq^a+C_2q^2\ .
\end{equation}
However, Yukawa's model is parity invariant, so that the term linear in $q^a$ cannot be present --the same is implied by the fact that the transition distributions must be symmetrical--. Being that way:
\begin{equation}\label{eq:4.2.31}
\widetilde\Pi(q)=\widehat\Pi(q)+C_0+C_2q^2\equiv \hat\Pi(q)+b+C_2(q^2-m_2^2)\ .
\end{equation}
In order to fix the normalization coefficients $b$ and $C_2$ we must impose some additional physical conditions. First, note that for $0<q^2<4m_1^2$ the number $\xi$ turns to be a complex number with unitary modulus, so it can be represented as:
\begin{equation}\label{eq:4.2.32}
\xi=e^{i\theta}\ ,\  q^2=4m_1^2\sin\left(\frac{\theta}{2}\right)^2\ .
\end{equation}
Therefore:
\begin{equation}\label{eq:4.2.33}
\frac{1+\xi}{1-\xi}\log(\xi)=2\sqrt{\frac{4m_1^2}{q^2}-1}\cot^{-1}\left(\sqrt{\frac{4m_1^2}{q^2}-1}\right)\ ,
\end{equation}
and the boson's self-energy takes the form:
\begin{align}
\widetilde\Pi(q)=&\frac{\mathfrak g^2}{(2\pi)^4}q^2\left\{\sqrt{\frac{4m_1^2}{q^2}-1}\cot^{-1}\left(\sqrt{\frac{4m_1^2}{q^2}-1}\right)+1\right\}\nonumber\\
&+b+C_2(q^2-m_2^2)\ .\label{eq:4.2.34}
\end{align}
On the other hand, in the case $q^2>4m_1^2$, substituting back $\xi$ from Eq. \eqref{eq:4.2.22} and using that $|\xi|=-\xi$ because in this case it is $\xi<0$:
\begin{align}
\widetilde \Pi(q)=&\frac{\mathfrak g^2}{2(2\pi)^4}q^2\Bigg\{\sqrt{1-\frac{4m_1^2}{q^2}}\log\left(\frac{1-\sqrt{1-4m_1^2/q^2}}{\sqrt{1-4m_1^2/q^2}+1}\right)\nonumber\\
&+2-i\pi\sqrt{1-\frac{4m_1^2}{q^2}}\Bigg\}+b+C_2(q^2-m_2^2)\ .\label{eq:4.2.35}
\end{align}

For the normalization we will study the fermion-fermion scattering with meson's self-energy insertions. Let us write the two-point distribution as:
\begin{equation}\label{eq:4.2.36}
T_2^{(FF)}(x_1;x_2)=i\mathfrak g^2\normord{j(x_1)t_2^{(FF)}(x_1-x_2)j(x_2)}\ ;
\end{equation}
\begin{equation*}
j(x)=\overline\psi(x)\gamma^5\psi(x)\ .
\end{equation*}
By the inductive procedure of CPT we construct the fourth order causal distribution for this process, which turns to be:
\begin{align}\label{eq:4.2.37}
D_4^{(FF)}&(x_1;x_2;x_3;x_4)=T_2^{(FF)}(x_1;x_3)T_2^{(FF)}(x_4;x_2)\nonumber\\
&\qquad-T_2^{(FF)}(x_2;x_4)T_2^{(FF)}(x_1;x_3)\nonumber\\
&=\mathfrak g^2\;\bm{\vcentcolon}\,j(x_1)t_2^{(FF)}(x_1-x_3)d(x_3-x_4)\nonumber\\
&\qquad\times t_2^{(FF)}(x_4-x_2)j(x_2)\,\bm{\vcentcolon}\;\ ,
\end{align}
with $d(y)$ the causal distribution for meson's self-energy given in Eq. \eqref{eq:4.2.2}. Since $t_2^{(FF)}$ has negative singular order it is sufficient to split the $d(y)$ distribution without obtaining divergences. In this manner:
\begin{align}\label{eq:4.2.38}
&T_4^{(FF)}(x_1;x_2;x_3;x_4)=-i\mathfrak g^2\;\bm{\vcentcolon}\,j(x_1)t_2^{(FF)}(x_1-x_3)\nonumber\\
&\qquad\times\Pi(x_3-x_4)t_2^{(FF)}(x_4-x_2)j(x_2)\,\bm{\vcentcolon}\;\ .
\end{align}
An analogous analysis holds for the next perturbation orders. Therefore, by defining the ``total meson propagator'' $D_{\text{tot}}$ by the sum of this series:
\begin{equation}\label{eq:4.2.39}
T^{(FF)}(x_1;x_2)=i\mathfrak g^2\normord{j(x_1)D_{\text{tot}}(x_1-x_2)j(x_2)}\ ,
\end{equation}
we will obtain that in momentum space it is given by:
\begin{align}
\widehat D_{\text{tot}}&=\hat t_2^{(FF)}-(2\pi)^4\hat t_2^{(FF)}\widetilde\Pi\hat t_2^{(FF)}\nonumber\\
&\quad+(2\pi)^8\hat t_2^{(FF)}\widetilde \Pi\hat t_2^{(FF)}\widetilde \Pi\hat t_2^{(FF)}-\cdots\nonumber\\
&=\hat t_2^{(FF)}\left(1-(2\pi)^4\widetilde\Pi\widehat D_{\text{tot}}\right)\ .\label{eq:4.2.40}
\end{align}
Since the two-point transition distribution is
\begin{equation}\label{eq:4.2.41}
\hat t_2^{(FF)}(p)=-(2\pi)^{-2}\frac{1}{p^2-m_2^2+i0^+}=\widehat D^F(p)\ ,
\end{equation}
we obtain that:
\begin{equation}\label{eq:4.2.42}
\widehat D_{\text{tot}}(q)=-(2\pi)^{-2}\frac{1}{q^2-\left(m_2^2+(2\pi)^2\widetilde\Pi(q)\right)+i0^+}\ .
\end{equation}
The physical conditions that we must impose are the following: (1) The physical mass of the pseudo-scalar particle is $m_2$, so $\widehat{D}_{\text{tot}}(q)$ must have its pole in that value, which occurs if:
\begin{equation}\label{eq:4.2.43}
\lim_{q^2\to m_2^2}\widetilde \Pi(q)=0\ ;
\end{equation}
(2) the physical value of the coupling constant is $g$. Since $\widehat D_{\text{tot}}(q)$ will multiply a current $g\overline\psi\gamma^5\psi$, the coefficient of $q^2$ in $\widehat D_{\text{tot}}(q)$ will divide effectively the value of $g$; then the normalization condition that we must impose is:
\begin{equation}\label{eq:4.2.44}
\lim_{q^2\to m_2^2}\frac{d\widetilde\Pi(q)}{dq^2}=0\ .
\end{equation}
Additionally, Yukawa's model, in the real world, describes the interaction between pions and nucleons, so the mass $m_2$ is the pion mass, which is less than the nucleon mass $m_1$. Because of this, the normalization conditions must be analysed with Eq. \eqref{eq:4.2.34}. Since there are no infra-red divergences, the limits in Eqs. \eqref{eq:4.2.43} and \eqref{eq:4.2.44} are equal to the simple evaluation at $q^2=m_2^2$. We obtain:
\begin{equation}\label{eq:4.2.45}
b=-\frac{\mathfrak g^2m_2^2}{2(2\pi)^{4}}\left\{2\sqrt{\frac{4m_1^2}{m_2^2}-1}\cot^{-1}\left(\sqrt{\frac{4m_1^2}{m_2^2}-1}\right)+2\right\}\ ,
\end{equation}
\begin{align}\label{eq:4.2.46}
C_2=&-\frac{\mathfrak g^2}{2(2\pi)^4}\Bigg\{\left(1-\frac{m_2^2}{4m_1^2-m_2^2}\right)\sqrt{\frac{4m_1^2}{m_2^2}-1}\nonumber\\
&\times\cot^{-1}\left(\sqrt{\frac{4m_1^2}{m_2^2}-1}\right)+3\Bigg\}\quad.
\end{align}
With this normalization, the obtained results are identical to those found in instant dynamics --see Ref. \cite{ABPS}--.

\section{Fermion's self-energy}

Now we turn to fermion's self-energy, which is described by the second order causal distribution \cite{APS2} --again, we use the relative coordinate $y=x_1-x_2$--:
\begin{align}
D_2^{(FSE)}&(x_1;x_2)=\mathfrak g^2\;\bm{\vcentcolon}\,\overline\psi(x_1)\gamma^5\big(S_+(y)D_+(y)\nonumber\\
&-S_-(y)D_-(y)\big)\gamma^5\psi(x_2)\,\bm{\vcentcolon}\;\nonumber\\
&-\mathfrak g^2\;\bm{\vcentcolon}\,\overline\psi(x_2)\gamma^5\big(S_+(-y)D_+(-y)\nonumber\\
&-S_-(-y)D_-(-y)\big)\gamma^5\psi(x_1)\,\bm{\vcentcolon}\;\ .\label{eq:4.3.1}
\end{align}
Defining the distributions:
\begin{equation}\label{eq:4.3.2}
d(y):=\mathfrak g^2\gamma^5\left(d_+(y)+d_-(y)\right)\gamma^5\ ,  
\end{equation}
\begin{equation*}
d_\pm(y):=\pm S_\pm(y)D_\pm(y)\ ,
\end{equation*}
Eq. \eqref{eq:4.3.1} can be written as:
\begin{align}
D_2^{(FSE)}(x_1;x_2)=&\normord{\overline\psi(x_1)d(y)\psi(x_2)}\nonumber\\
&-\normord{\overline\psi(x_2)d(-y)\psi(x_1)}\ .\label{eq:4.3.3}
\end{align}
As we can see, the second term can be obtained from the first one by the exchange of $x_1$ and $x_2$, so we only need to focus on the first term. Again, no instantaneous term (no spurious pole) appears here due to the implementation of the causal inductive procedure. We apply Fourier's transformation in order to go to momentum space. Starting with $d_-$:
\begin{align}
&\hat d_-(p)=-(2\pi)^{-2}\int d^4y e^{ipy}S_-(y)D_-(y)\nonumber\\
&\quad=-(2\pi)^{-6}\int d^4yd^4qd^4ke^{i(p-q-k)y}\widehat S_-(q)\widehat D_-(k)\ .\label{eq:4.3.4}
\end{align}
Integrating in the variable $y$, then in the variable $k$:
\begin{equation}\label{eq:4.3.5}
\hat d_-(p)=-(2\pi)^{-2}\int d^4q\widehat S_-(q)\widehat D_-(p-q)\ .
\end{equation}
But:
\begin{equation}\label{eq:4.3.6}
\widehat S_-(q)=(\slashed q+m_1)\widehat D_{-,m_1}(q)\ ,
\end{equation}
\begin{equation}\label{eq:4.3.6-1}
\widehat D_{-,m}(q)=-\frac{i}{2\pi}\Theta\left(-q_-\right)\delta(q^2-m^2)\ ,
\end{equation}
which, substituted into Eq. \eqref{eq:4.3.5}, leads us to the following expression:
\begin{equation}\label{eq:4.3.7}
\hat d_-(p)=(2\pi)^{-4}\left[\gamma^a I_{2a}(p)+m_1I_1(p)\right]\ ,
\end{equation}
with the integrals:
\begin{align}\label{eq:4.3.8}
I_1(p)=&\int d^4q\Theta\left(-q_-\right)\Theta\left(q_--p_-\right)\delta(q^2-m_1^2)\nonumber\\
&\times\delta\left((p-q)^2-m_2^2\right)\ ,
\end{align}
\begin{align}\label{eq:4.3.9}
I_{2a}(p)=&\int d^4qq_a\Theta\left(-q_-\right)\Theta\left(q_--p_-\right)\delta(q^2-m_1^2)\nonumber\\
&\times\delta\left((p-q)^2-m_2^2\right)\ .
\end{align}
As usual, the evaluation of these integrals is simplified in a convenient reference frame, such that $(p_a)=\left(p_+;0_\perp;p_-\right)$; such a reference frame exists because, as implied by the supports of Dirac's deltas distributions and Heaviside's functions in Eqs. \eqref{eq:4.3.8} and \eqref{eq:4.3.9}, $(p-q)\in V^-(0)$ and $q\in V^-(0)$, so $p=(p-q)+q\in V^-(0)$ as well. Clearly, this implies, in particular, that $p_+,\ p_-<0$. In this reference frame, Eqs. \eqref{eq:4.3.8} and \eqref{eq:4.3.9} are:
\begin{align}
&I_1=\int d^4q\Theta\left(-q_-\right)\Theta\left(q_--p_-\right)\delta(q^2-m_1^2)\nonumber\\
&\quad\times\delta\left(2p_+p_--2p_-q_+-2p_+q_-+(m_1^2-m_2^2)\right)\ ,\label{eq:4.3.10}\\
&I_{2\pm}=\int d^4qq_{\pm}\Theta\left(-q_-\right)\Theta\left(q_--p_-\right)\delta(q^2-m_1^2)\nonumber\\
&\quad\times\delta\left(2p_+p_--2p_-q_+-2p_+q_-+(m_1^2-m_2^2)\right)\ ,\label{eq:4.3.11}\\
&I_{2\perp}=0\ .\label{eq:4.3.12}
\end{align}
Particularly, Eq. \eqref{eq:4.3.12} comes from the fact that the integrand in Eq. \eqref{eq:4.3.9} becomes odd for $q_a=q_\perp$ in the chosen reference frame. Also, the product of two Dirac's delta distributions appearing in these integrals can be put in the following form by using their properties:
\begin{align}
&\frac{1}{|8p_-|}\Theta\left(A^2-\frac{p_+\omega_q^2}{2p_-}\right)\delta\left(q_--\frac{\omega_q^2}{2q_+}\right)F(q_+;q_\perp)\ ,\nonumber
\end{align}
with the quantities $F(q_+;q_\perp)$ and $A$ defined as:
\begin{align}
&F(q_+;q_\perp)=\frac{1}{\sqrt{A^2-\dfrac{p_+\omega_q^2}{2p_-}}}\nonumber\\
&\quad\times\left\{\delta\left[q_+-\left(A+\sqrt{A^2-\frac{p_+\omega_q^2}{2p_-}}\right)\right]\right.\nonumber\\
&\quad\quad\left.+\delta\left[q_+-\left(A-\sqrt{A^2-\frac{p_+\omega_q^2}{2p_-}}\right)\right]\right\}\ ,\label{eq:4.3.13}
\end{align}
\begin{equation}\label{eq:4.3.14}
 A=\frac{2p_+p_-+(m_1^2-m_2^2)}{4p_-}\ .
\end{equation}
This quantity $A$ is constant in the integration process. Now, since $\omega_q^2=q_\perp^2+m_1^2>m_1^2$, Heaviside's function in Eq. \eqref{eq:4.3.13} implies that: $2p_-A^2/p_+>m_1^2$. Putting the value of $A$ in this inequality we arrive at:
\begin{equation}\label{eq:4.3.15}
 \left(2p_+p_--(m_1^2+m_2^2)\right)^2-4m_1^2m_2^2>0\ .
\end{equation}
With these manipulations, and writing $d^2q_\perp=\pi d(\omega_q^2)$, the integrals in Eqs. \eqref{eq:4.3.10} and \eqref{eq:4.3.11} are:
\begin{align}
&I_1=\frac{\pi\Theta(-p_-)}{8|p_-|}\Theta\left(\left(2p_+p_--(m_1^2+m_2^2)\right)^2-4m_1^2m_2^2\right)\nonumber\\
&\quad\times\int\limits_{m_1^2}^{2p_-A^2/p_+}d(\omega_q^2)\int\limits_{\omega_q^2/2p_-}^{0}dq_+F(q_+;q_\perp)\ ,\label{eq:4.3.16}
\end{align}
\begin{align}
&I_{2+}=\frac{\pi\Theta(-p_-)}{8|p_-|}\Theta\left(\left(2p_+p_--(m_1^2+m_2^2)\right)^2-4m_1^2m_2^2\right)\nonumber\\
&\quad\times\int\limits_{m_1^2}^{2p_-A^2/p_+}d(\omega_q^2)\int\limits_{\omega_q^2/2p_-}^{0}dq_+\ q_+F(q_+;q_\perp)\ ,\label{eq:4.3.17}
\end{align}
\begin{align}
&I_{2-}=\frac{\pi\Theta(-p_-)}{8|p_-|}\Theta\left(\left(2p_+p_--(m_1^2+m_2^2)\right)^2-4m_1^2m_2^2\right)\nonumber\\
&\quad\times\int\limits_{m_1^2}^{2p_-A^2/p_+}d(\omega_q^2)\int\limits_{\omega_q^2/2p_-}^{0}dq_+\frac{\omega_q^2}{2q_+}F(q_+;q_\perp)\ .\label{eq:4.3.18}
\end{align}
In order to integrate in the variable $q_+$ we need to see under what conditions the Dirac's delta distributions are non-null. Since $q_+<0$, it must be $A<0$, which means [see Eq. \eqref{eq:4.3.14} and remember that $p_-<0$]:
\begin{equation}\label{eq:4.3.19}
2p_+p_->m_1^2-m_2^2\ .
\end{equation}
Additionally, the argument of Heaviside's functions in Eqs. \eqref{eq:4.3.16}-\eqref{eq:4.3.18} can be written as:
\begin{equation}\label{eq:4.3.20}
\left(2p_+p_--(m_1^2-m_2^2)\right)^2-8p_+p_-m_2^2>0\ ,
\end{equation}
which jointly with Eq. \eqref{eq:4.3.19} leads to the inequality: $2p_+p_--2m_2\sqrt{2p_+p_-}-(m_1^2-m^2_2)>0$, with roots $\sqrt{2p_+p_-}=m_2\pm m_1$. Under the assumption that $m_1>m_2$ --for example, the mass of the nucleons is greater than that of the pions: $m_N\approx 940$ MeV/c$^2$, and $m_{\pi}\approx 140$ MeV/c$^2$--, we find that the integration in the variable $q_+$ is proportional to:
\begin{equation}\label{eq:4.3.21}
\Theta\left[2p_+p_--(m_1+m_2)^2\right]\ .
\end{equation}
For the integration in the variable $\omega_q^2$, we use the following result:
\begin{align}
&\int\limits_{m_1^2}^{2p_-A^2/p_+}\frac{d(\omega_q^2)}{\sqrt{A^2-\dfrac{p_+\omega_q^2}{2p_-}}}=4\sqrt{\frac{p_-}{p_+}}\sqrt{\frac{p_-A^2}{p_+}-\frac{m_1^2}{2}}\nonumber\\
&\qquad=2|p_-|\sqrt{1-\frac{2(m_1^2+m_2^2)}{2p_+p_-}+\frac{(m_1^2-m_2^2)^2}{(2p_+p_-)^2}}\ .\label{eq:4.3.22}
\end{align}
The required integrals are then, in Lorentz's covariant form:
\begin{align}
&I_1=\frac{\pi}{2}\Theta\left(-p_-\right)\Theta\left[p^2-(m_1+m_2)^2\right]\nonumber\\
&\quad\times\sqrt{1-\frac{2(m_1^2+m_2^2)}{p^2}+\frac{(m_1^2-m_2^2)^2}{p^4}}\ ,\label{eq:4.3.25}\\
&I_{2a}=\frac{p_a}{2}\left(1+\frac{m_1^2-m_2^2}{p^2}\right)I_1\ .\label{eq:4.3.26}
\end{align}
Substituting Eqs. \eqref{eq:4.3.25} and \eqref{eq:4.3.26} into Eq. \eqref{eq:4.3.7}, and using that $\left(\gamma^5\right)^2=1$ and $\gamma^5\gamma^a\gamma^5=-\gamma^a$:
\begin{align}
\mathfrak g^2\gamma^5&\hat d_-(p)\gamma^5=\frac{\mathfrak g^2}{4}(2\pi)^{-3}\Theta\left(-p_-\right)\Theta\left[p^2-(m_1+m_2)^2\right]\nonumber\\
&\times\sqrt{1-\frac{2(m_1^2+m_2^2)}{p^2}+\frac{(m_1^2-m_2^2)^2}{p^4}}\nonumber\\
&\times\left\{m_1-\frac{\slashed p}{2}\left(1+\frac{m_1^2-m_2^2}{p^2}\right)\right\}\ .\label{eq:4.3.27}
\end{align}

For the distribution $d_+$ we must follow the same steps; the result is:
\begin{align}
\mathfrak g^2\gamma^5&\hat d_+(p)\gamma^5=-\frac{\mathfrak g^2}{4}(2\pi)^{-3}\Theta\left(p_-\right)\Theta\left[p^2-(m_1+m_2)^2\right]\nonumber\\
&\times\sqrt{1-\frac{2(m_1^2+m_2^2)}{p^2}+\frac{(m_1^2-m_2^2)^2}{p^4}}\nonumber\\
&\times\left\{m_1-\frac{\slashed p}{2}\left(1+\frac{m_1^2-m_2^2}{p^2}\right)\right\}\ .\label{eq:4.3.28}
\end{align}
Being that way, the numerical part of the causal distribution describing fermion's self-energy, in momentum space, is [see Eq. \eqref{eq:4.3.2}]:
\begin{align}
\hat d(p)=&-\frac{\mathfrak g^2}{4}(2\pi)^{-3}\text{sgn}\left(p_-\right)\Theta\left[p^2-(m_1+m_2)^2\right]\nonumber\\
&\times\sqrt{1-\frac{2(m_1^2+m_2^2)}{p^2}+\frac{(m_1^2-m_2^2)^2}{p^4}}\nonumber\\
&\times\left\{m_1-\frac{\slashed p}{2}\left(1+\frac{m_1^2-m_2^2}{p^2}\right)\right\}\ .\label{eq:4.3.29}
\end{align}

In order to split this causal distribution and obtain its retarded part we will write it in a convenient way by factorizing a polynomial:
\begin{equation}\label{eq:4.3.30}
\hat d(p)=-\frac{\mathfrak g^2}{4}(2\pi)^{-3}\left\{m_1p^2-\frac{\slashed p}{2}\left[p^2+(m_1^2-m_2^2)\right]\right\}\hat d_1(p)\ ,
\end{equation}
with:
\begin{align}
\hat d_1(p)=&\text{sgn}(p_-)\Theta\left[p^2-(m_1+m_2)^2\right]\nonumber\\
&\times\frac{1}{p^2}\sqrt{1-\frac{2(m_1^2+m_2^2)}{p^2}+\frac{(m_1^2-m_2^2)^2}{p^4}}\ .\label{eq:4.3.31}
\end{align}
Then, by theorem it is sufficient to split the distribution $\hat d_1(p)$. Going to a reference frame in which $(p_a)=(p_+;0_\perp;p_-)$, which is possible since $p\in V^+$ by the support of Dirac's delta distribution and Heaviside's function in Eq. \eqref{eq:4.3.32}:
\begin{align}
&\hat d_1(p)=\text{sgn}(p_-)\Theta\left[2p_+p_--(m_1+m_2)^2\right]\nonumber\\
&\quad\times\frac{1}{2p_+p_-}\sqrt{1-\frac{2(m_1^2+m_2^2)}{2p_+p_-}+\frac{(m_1^2-m_2^2)^2}{(2p_+p_-)^2}}\ .\label{eq:4.3.32}
\end{align}
This distribution has singular order $\omega_-^1=-1<0$, so its retarded part is simply given by:
\begin{align}
\hat r_1(p)&=\frac{i}{2\pi}\int \frac{dk}{k+i0^+}\text{sgn}(p_-)\frac{1}{-2kp_-+2p_+p_-}\nonumber\\
&\times\Theta\left[-2kp_-+2p_+p_--(m_1+m_2)^2\right]\nonumber\\
&\times\sqrt{1-\frac{2(m_1^2+m_2^2)}{-2kp_-+2p_+p_-}+\frac{(m_1^2-m_2^2)^2}{(-2kp_-+2p_+p_-)^2}}\ .\label{eq:4.3.33}
\end{align}
Using Sokhotskiy's formula for treating the pole, then making the change $s=-2kp_-+2p_+p_-$:
\begin{align}
\hat r_{1}(p)=&\frac{i}{2\pi}\int\limits_{(m_1+m_2)^2}^{+\infty}\frac{ds}{s(2p_+p_--s)}\nonumber\\
&\times\sqrt{1-\frac{2(m_1^2+m_2^2)}{s}+\frac{(m_1^2-m_2^2)^2}{s^2}}\nonumber\\
&+\frac{1}{2}\text{sgn}\left(p_-\right)\Theta(2p_+p_--(m_1+m_2)^2)\frac{1}{2p_+p_-}\nonumber\\
&\times\sqrt{1-\frac{2(m_1^2+m_2^2)}{2p_+p_-}+\frac{(m_1^2-m_2^2)^2}{(2p_+p_-)^2}}\ .\label{eq:4.3.34}
\end{align}
Let us call $J$ the integral in Eq. \eqref{eq:4.3.34}. It can be written as:
\begin{align}\label{eq:4.3.35}
J=&\int\limits_{(m_1+m_2)^2}^{+\infty}\frac{ds}{s^2(2p_+p_--s)}\nonumber\\
&\times\sqrt{[s-(m_1+m_2)^2][s-(m_1-m_2)^2]}\ .
\end{align}
This integral can be solved by the third Euler's substitution \cite{Piskunov}: We perform the change of variable:
\begin{equation}\label{eq:4.3.36}
s=\frac{(m_1+m_2)^2-(m_1-m_2)^2x^2}{1-x^2}\ ;\  0<x<1\ ,
\end{equation}
with which:
\begin{equation*}
\sqrt{[s-(m_1+m_2)^2][s-(m_1-m_2)^2]}=4m_1m_2\frac{x}{1-x^2}\ ,
\end{equation*}
\begin{equation*}
ds=\frac{8m_1m_2xdx}{(1+x)^2(1-x)^2}\ .
\end{equation*}
Defining also the parameters $a$ and $b$ according to:
\begin{equation}\label{eq:4.3.38}
a^2=\frac{(m_1+m_2)^2}{(m_1-m_2)^2}>1\ ,\  b^2=\left|\frac{2p_+p_--(m_1+m_2)^2}{2p_+p_--(m_1-m_2)^2}\right|\ ,
\end{equation}
the integral $J$ adopts the form:
\begin{align}
J_\eta=&-\frac{8m_1m_2(a^2-b^2)}{2p_+p_-(m_1-m_2)^2}\nonumber\\
&\times\int\limits_0^1\frac{x^2dx}{(a+x)^2(a-x)^2(x^2-\eta b^2)}\ ,\label{eq:4.3.39}
\end{align}
with $\eta=+1$ if $2p_+p_-<(m_1-m_2)^2$ or $2p_+p_->(m_1+m_2)^2$ and $\eta=-1$ if $(m_1-m_2)^2<2p_+p_-<(m_1+m_2)^2$. This integrals can already be evaluated by the technique of partial fractions. We obtain the following results:
\begin{align}
J_{+1}=&\frac{2p_+p_--(m_1-m_2)^2}{(2p_+p_-)^2}\left\{\frac{b^2-a^2}{a^2-1}+b\log\left(\left|\frac{1+b}{1-b}\right|\right)\right.\nonumber\\
&\left.-\frac{a^2+b^2}{2a}\log\left(\frac{a+1}{a-1}\right)\right\}\ ;\label{eq:4.3.40}
\end{align}
\begin{align}
J_{-1}=&\frac{2p_+p_--(m_1-m_2)^2}{(2p_+p_-)^2}\frac{(b^2-a^2)^2}{2a(a^2+b^2)^2}\nonumber\\
&\times\left\{(b^2-a^2)\log\left(\frac{a+1}{a-1}\right)\right.\nonumber\\
&\left.+4ab\tan^{-1}\left(\frac{1}{b}\right)-\frac{2a(a^2+b^2)}{a^2-1}\right\}\ .\label{eq:4.3.40.1}
\end{align}

The Lorentz's covariant version of this distribution is obtained by the replacement of $2p_+p_-$ by $p^2$. By virtue of Eqs. \eqref{eq:4.3.30} and \eqref{eq:4.3.34} the retarded distribution is:
\begin{align}
&\hat r(p)=-\frac{i\mathfrak g^2}{4(2\pi)^4}\Bigg\{\left(m_1p^2-\frac{\slashed p}{2}\left[p^2+(m_1^2-m_2^2)\right]\right)J_\eta(p)\nonumber\\
&\quad-i\pi\text{sgn}\left(p_-\right)\Theta\left(p^2-(m_1+m_2)^2\right)\nonumber\\
&\quad\times\left[m_1-\frac{\slashed p}{2}\left(1+\frac{m_1^2-m_2^2}{p^2}\right)\right]\nonumber\\
&\quad\times\sqrt{1-\frac{2(m_1^2+m_2^2)}{p^2}+\frac{(m_1^2-m_2^2)^2}{p^4}}\Bigg\}\ .\label{eq:4.3.41}
\end{align}
The numerical distribution $\hat t(p)$ contained in the transition distribution is finally obtained by subtracting the subsidiary distribution $\hat r'(p)$: $\hat t(p)=\hat r(p)-\hat r'(p)$. Therefore, defining the ``fermion's self-energy'' $\widehat \Sigma(p)$ by:
\begin{equation*}
\hat t(p)=:-i\hat\Sigma(p)\ ;
\end{equation*}
\begin{equation}\label{eq:4.3.42}
T_2^{(FSE)}(x_1;x_2)=-i\normord{\overline\psi(x_1)\Sigma(x_1-x_2)\psi(x_2)}\ ,
\end{equation}
from Eq. \eqref{eq:4.3.41} it follows that:
\begin{align}
&\widehat \Sigma(p)=\frac{\mathfrak g^2}{4(2\pi)^4}\Bigg\{\left(m_1p^2-\frac{\slashed p}{2}\left[p^2+(m_1^2-m_2^2)\right]\right)J_\eta(p)\nonumber\\
&\quad-i\pi\Theta\left(p^2-(m_1+m_2)^2\right)\left[m_1-\frac{\slashed p}{2}\left(1+\frac{m_1^2-m_2^2}{p^2}\right)\right]\nonumber\\
&\quad\times\sqrt{1-\frac{2(m_1^2+m_2^2)}{p^2}+\frac{(m_1^2-m_2^2)^2}{p^4}}\Bigg\}\ .\label{eq:4.3.43}
\end{align}

The singular order of $\widehat\Sigma(p)$ at the $x^-$-axis is $\omega_-=+1$, hence its most general form is:
\begin{equation}\label{eq:4.3.44}
\widetilde\Sigma(p)=\widehat\Sigma(p)+C_0+C_1\slashed p\equiv \widehat\Sigma(p)+c+C_1(\slashed p-m_1)\ .
\end{equation}
As in the case of boson's self-energy, the constants $c$ and $C_1$ will be fixed by imposing additional physical conditions. Those could be imposed to the total fermion propagator, defined as the series for the meson-fermion scattering:
\begin{align}\label{eq:4.3.45}
T^{(BF)}(x_1;x_2)=&-i\mathfrak g^2\normord{\overline\psi(x_1)\gamma^5S_{\text{tot}}(x_1-x_2)\gamma^5\psi(x_2)}\nonumber\\
&\times\normord{\varphi(x_1)\varphi(x_2)}\ .
\end{align}
To calculate the fourth order contribution to this series we start, in the inductive process, by constructing the causal distribution:
\begin{align}
D_4^{(BF)}&(x_1;x_2;x_3;x_4)=T_2^{(BF)}(x_1;x_3)T_2^{(BF)}(x_4;x_2)\nonumber\\
&-T_2^{(BF)}(x_2;x_4)T_2^{(BF)}(x_1;x_3)\ .\label{eq:4.3.46}
\end{align}
By writing the second order transition distribution for the meson-fermion scattering as:
\begin{align}\label{eq:4.3.47}
T_2^{(BF)}(x_1;x_2)=&i\mathfrak g^2\normord{\overline\psi(x_1)\gamma^5 t_2^{(BF)}(x_1-x_2)\gamma^5\psi(x_2)}\nonumber\\
&\times\normord{\varphi(x_1)\varphi(x_2)}\ ,
\end{align}
we obtain by using Wick's theorem:
\begin{align}
&D_4^{(BF)}(x_1;x_2;x_3;x_4)=\mathfrak g^2\;\bm{\vcentcolon}\,\overline\psi(x_1)\gamma^5t_2^{(BF)}(x_1-x_3)\nonumber\\
&\quad\times d(x_3-x_4)t_2^{(BF)}(x_4-x_2)\psi(x_2)\,\bm{\vcentcolon}\;\normord{\varphi(x_1)\varphi(x_2)}\ ,\label{eq:4.3.48}
\end{align}
and its corresponding transition distribution can be obtained by the splitting of the distribution $d(x_3-x_4)$, which is the one in Eq. \eqref{eq:4.3.2}, because the distributions $t_2^{(BF)}$ have negative singular order:
\begin{align}
&T_4^{(BF)}(x_1;x_2;x_3;x_4)=-i\mathfrak g^2\;\bm{\vcentcolon}\,\overline\psi(x_1)\gamma^5 t_1^{(BF)}(x_1-x_3)\nonumber\\
&\quad\times\Sigma(x_3-x_4)t_2^{(BF)}(x_4-x_2)\psi(x_2)\,\bm{\vcentcolon}\;\normord{\varphi(x_1)\varphi(x_2)}\ ,\label{eq:4.3.49}
\end{align}
and similarly for the next order perturbation terms. As a consequence, the total fermion propagator in momentum space is given by the series:
\begin{align}
\widehat S_{\text{tot}}&=-\hat t_2^{(BF)}+(2\pi)^4\hat t_2^{(BF)}\widetilde\Sigma\hat t_2^{(BF)}\nonumber\\
&\quad-(2\pi)^8\hat t_2^{(BF)}\widetilde \Sigma\hat t_2^{(BF)}\widetilde\Sigma\hat t_2^{(BF)}+\cdots\nonumber\\
&=-\hat t_2^{(BF)}\left(1+(2\pi)^4\widetilde\Sigma\widehat S_{\text{tot}}\right)\ .\label{eq:4.3.50}
\end{align}
Now, the two-point transition distribution for the meson-fermion scattering is given by:
\begin{align}
\hat t_2^{(BF)}(p)&=-(2\pi)^{-2}\frac{\slashed p+m_1}{p^2-m_1^2+i0^+}\nonumber\\
&=-(2\pi)^{-2}\frac{1}{\slashed p-m_1+i0^+}\ ,\label{eq:4.3.51}
\end{align}
because the distributions of a given order are constructed with the already normalized transition distributions of less order, as explained in Subsec. \ref{subsec:I.A}. In this case, although Feynman's propagator of the fermion field has an instantaneous term, it is eliminated by a suitable normalization at tree level and no more appears in the next-order calculations --for details, see Ref. \cite{APS2} --. Accordingly, from Eq. \eqref{eq:4.3.50} it follows that:
\begin{equation}\label{eq:4.3.52}
\widehat S_{\text{tot}}(p)=(2\pi)^{-2}\frac{1}{\slashed p-\left(m_1+(2\pi)^2\widetilde\Sigma(p)\right)+i0^+}\ .
\end{equation}
The physical conditions which this propagator must satisfy are: (1) The physical mass of the fermion is $m_1$, so $\widehat S_{\text{tot}}(p)$ must have a pole in $\slashed p=m_1$; (2) the physical value of the coupling constant is $\mathfrak g$, hence the coefficient of $\slashed p$ in the denominator of $\widehat S_{\text{tot}}(p)$ must be one. In this way:
\begin{equation}\label{eq:4.3.53}
\lim_{\slashed p\to m_1}\widetilde\Sigma(p)=0\ \text{and}\ \lim_{\slashed p\to m_1}\frac{d\widetilde{\Sigma}(p)}{d\slashed p}=0\ .
\end{equation}
These two conditions are satisfied by the choice --we use the case with $\eta=-1$ because it is the one which contains the mass-shell $p^2=m_1^2$--:
\begin{equation*}
c=-\frac{\mathfrak g^2m_1m_2^2}{8(2\pi)^4}J_{-1}(\slashed p=m_1)\ ,
\end{equation*}
\begin{equation*}
C_1=-\frac{\mathfrak g^2m_2^2}{8(2\pi)^4}\left\{J_{-1}(\slashed p=m_1)+m_1\frac{dJ_{-1}}{d\slashed p}(\slashed p=m_1)\right\}\ .
\end{equation*}
As in meson's case, also fermion's self-energy in light-front dynamics is equal to the result obtained in instant dynamics \cite{ABPS}.

\section{Conclusions}

We have obtained the expression for boson's and fermion's self-energies in the neutral Yukawa's model in light-front dynamics without ambiguities and avoiding the complications of the different regularization schemes. In particular, null-plane CPT has shown to be very useful because, since it does not use Feynman's rules but the causal inductive procedure, no Feynman's propagator appears as part of ``loop distributions'', avoiding the appearance of the spurious pole that the fermion Feynman's propagator contains in its instantaneous term, and whose removal is a major problem in the usual approaches.

Our results must be compared with those obtained in instant dynamics \cite{ABPS}, which shows in a very direct manner the equivalence of both dynamical forms.

\begin{acknowledgments}
 O.A.A. and B.M.P. thank CNPq-Brazil for total and partial support, respectively. The authors also thank the anonymous referees for  the useful observations and suggestions which helped us to improve our exposition.
\end{acknowledgments}

\bibliography{AcevedoPimentel}

\end{document}